\documentstyle[amstex,amsfonts,epsfig,supercite]{article}

\begin{document}
\begin{center}
{\LARGE \bf The Osmotic Coefficient of Rod-like Polyelectrolytes:\\[0.2cm]
 Computer
  Simulation, Analytical \\[0.2cm] Theory, and Experiment}\\[0.5cm]

{\it \large  M. Deserno$^1$, C. Holm$^1$, J. Blaul$^2$, M. Ballauff$^2$, and M. Rehahn$^3$}\\[0.2cm]


{$^1$ Max-Planck-Institut f\"ur Polymerforschung, Ackermannweg 10,
  55128 Mainz, Germany\\
  $^2$ Polymer-Institut, Universit\"at Karlsruhe, Kaiserstra{\ss}e 12, 76128
  Karlsruhe, Germany\\
  $^3$ Institut f\"ur Makromolekulare Chemie, Technische Universit\"at
  Darmstadt,
  Petersenstra{\ss}e 22, 64287 Darmstadt, Germany}\\[0.5cm]
\end{center}
\date{\today}



\newcommand{\D}{\displaystyle}

\newcommand{\MD}{}

\newcommand{\romB}{{\operatorname{B}}}
\newcommand{\romLJ}{{\operatorname{LJ}}}
\newcommand{\romM}{{\operatorname{M}}}
\newcommand{\romPB}{{\operatorname{PB}}}

\newcommand{\romb}{{\operatorname{b}}}
\newcommand{\romc}{{\operatorname{c}}}
\newcommand{\romcut}{{\operatorname{cut}}}
\newcommand{\romid}{{\operatorname{id}}}
\newcommand{\romp}{{\operatorname{p}}}
\newcommand{\romr}{{\operatorname{r}}}
\newcommand{\roms}{{\operatorname{s}}}


\begin{abstract}
  The osmotic coefficient of solutions of rod-like polyelectrolytes is
  considered by comparing current theoretical treatments and
  simulations to recent experimental data.  The discussion is
  restricted to the case of monovalent counterions and dilute,
  salt-free solutions.  {\MD The classical Poisson-Boltzmann solution
  of the cell model correctly predicts a strong decrease in the
  osmotic coefficient, but upon closer look systematically
  overestimates its value.  The contribution of ion-ion-correlations
  are quantitatively studied by MD simulations and the recently
  proposed DHHC theory. However, our comparison with experimental data
  obtained on synthetic, stiff-chain polyelectrolytes shows that
  correlation effects can only partly explain the discrepancy.  A
  quantitative understanding thus requires theoretical efforts beyond
  the restricted primitive model of electrolytes.}
\end{abstract}




\section{Introduction}

Polyelectrolytes are macromolecules carrying charged groups which in polar
solvents can dissociate into a highly charged macroion and oppositely charged
small counterions \cite{Man88,Oos71,Kat71,BaJo96,Sch93}. The high electric
field of the macroion strongly couples to the counterions, which in turn will
tend to {\MD partly} neutralize the macroion. This phenomenon has been termed
``counterion condensation'' \cite{Man69,Man72,Man74}. Since polyelectrolytes
present an ubiquitous class of materials, the quantitative understanding of
counterion condensation is of fundamental importance.

The strong correlation of the counterions with the macroion reduces their
thermodynamic activity. Only a certain fraction of ions will be osmotically
active and contribute to the osmotic pressure $\Pi$ of polyelectrolytes in
dilute solutions, as Kern was able to show by osmotic measurements as early as
1938 \cite{Ker38}. The reduced activity can be expressed in terms of the
osmotic coefficient
\begin{equation}
  \phi := \frac{\Pi}{\Pi_\romid},
\end{equation}
where $\Pi_\romid$ is the ideal osmotic pressure calculated for a solution of
counterions {\MD interacting neither with themselves nor with the macroion}. A
great number of experimental studies of $\phi$ for strongly charged
polyelectrolytes is now available in literature, demonstrating that $\phi$ is
of the order of $0.2$ to $0.3$ for univalent counterions in the dilute regime
\cite{Man88,Oos71,Kat71,AuAl69,ReMa70,TaKa70,KoKr71,Dol74,KaYa98,OpWa99,BlWi}.
Furthermore, a comparison of experimental data obtained for different
counterions show that specific interactions between counterions and the
macroion can lead to an additional reduction of the measured osmotic pressure
\cite{Man88,Kat71}. In case of monovalent counterions, these specific effects
are small as compared to the effect of counterion condensation that presents
the dominant feature of strongly charged polyelectrolytes. A detailed
discussion of this problem has been given by Mandel \cite{Man88}. However, no
quantitative conclusions could be drawn from this comparison so far, since
practically all experimental data have been obtained from solutions of
flexible polyelectrolytes, which change their conformation upon lowering the
ionic strength \cite{ste95a}. The intricate connection of the osmotic
coefficient with the conformation of the macroion is not yet fully understood,
therefore studies of flexible polyelectrolytes in solution cannot lead to
direct conclusions on the validity of theories developed for rod-like
macroions.

In this paper we wish to discuss recent theoretical models of rod-like
polyelectrolytes and compare the calculated osmotic coefficient to results of
experimental investigations of a suitable model system. We restrict the
discussion to the case of rod-like macroions for which a partial counterion
condensation occurs \cite{Man69,FuKa51,AlBe51,BrZi84}, i.e., part of the
counterions remain within a finite distance to the macroion and will not be
diluted away when lowering the concentration to extremely low values. The
paper is organized as follows: First the simplified situation of the
cylindrical cell model and its solution within the framework of
Poisson-Boltzmann (PB) theory will be discussed. Next we consider theoretical
refinements designed to overcome limitations inherent to the cell model or its
mean-field PB solution. In particular, computer simulations will be used to
assess the importance of correlations neglected in the PB approach. Finally
recent measurements of the osmotic coefficient for a synthetic rod-like
polyelectrolyte will be quantitatively compared to two theoretical models and
computer simulations.  Hence, the present comparison allows to assess to what
extend the deviations between measurements and PB theory can be accounted for
by the modern theoretical approaches.


\section{The cell model}

The cell model is a commonly used way of reducing the complicated many-body
problem of a polyelectrolyte solution to an effective one-particle theory,
i.e., the case of a single polyelectrolyte chain in a cell
\cite{Kat71,BaJo96,Sch93,FuKa51,AlBe51,Man92}. The idea is to partition the
solution into sub-volumes, each containing only a single macroion together
with its counterions.  Since each sub-volume is electrically neutral, the
electric field will {\em on average\/} vanish on the cell surface. By virtue
of this construction different sub-volumes are electrostatically decoupled to
a first approximation. One may thus hope to factorize the partition function
and reduce the problem to the treatment of just one sub-volume, called
``cell''.  Its shape is assumed to reflect the symmetry of the
polyelectrolyte.  For a solution of rod-like polyelectrolytes with density
$\rho_\romp$ and rod length $L$ this gives a cylindrical cell with the radius
$R$ being fixed by the condition $\pi R^2 L\times\rho_\romp=1$. The
theoretical treatment is much simpler after neg\-lec\-ting end effects at the
cylinder caps. This is equivalent to taking $L=\infty$ {\em after\/} mapping
to the correct density.

The ions will be described as point-like, but they will electrostatically
interact with the macroion as well as with each other.  Their positions may
thus be strongly correlated. The solvent molecules are not explicitly taken
into account; rather, they are assumed to form a continuous dielectric
background which is completely specified by its dielectric constant
$\varepsilon_\romr$.

In this simplified situation it can be proved rigorously that the osmotic
pressure is given by the counterion density at the cell boundary times the
thermal energy $k_\romB T$ \cite{WeJo82}. The osmotic coefficient is thus
given by the ratio of the ion density at the cell boundary divided by the
average ion density.


\section{Poisson-Boltzmann theory for the cylindrical cell model}

The analytical description taking into account all individual counterions
turns out to be still too involved, since -- {\MD as usual} -- inter-ionic
correlations complicate matters. The {\MD standard} way out is neglecting
those correlations in a mean-field spirit, as is done in the Poisson-Boltzmann
description: The ionic degrees of freedom are replaced by a cylindrical
counterion density, and their interaction is approximated by the assumption
that the density is locally proportional to the Boltzmann-factor, see e.g.
Refs.~\cite{FuKa51,AlBe51,BrZi84,DeHo00a}. It can be shown that on this
mean-field level the osmotic coefficient is still given by the ratio between
boundary density and average density \cite{Mar55}.

Let us briefly recapitulate the Poisson-Boltzmann treatment of the cell model.
Consider a rod of radius $r_0$ and line charge density $\lambda>0$ coaxially
enclosed in a cylinder of radius $R$. Electroneutrality is achieved by adding
the appropriate amount of monovalent counterions; no additional salt shall be
present. The strength of the electrostatic interactions is conveniently
expressed by the Bjerrum length
\begin{equation}
  \ell_\romB = \frac{e_0^2}{4\pi\varepsilon_0\varepsilon_\romr k_\romB T},
\end{equation}
where $e_0$ is the unit charge, $\varepsilon_\romr$ is the dielectric
constant of the solvent, and $\varepsilon_0$, $k_\romB$ and $T$ have
their usual meaning. This definition suggests a dimensionless way of
measuring the line charge density of the rod via the charge parameter
$\xi$:
\begin{equation}
  \xi = \lambda \ell_\romB / e_0.
\end{equation}
It counts the number of unit charges on the rod per Bjerrum length and
is usually called the Manning parameter. In the following only the
strongly charged case $\xi>1$ will be discussed.

Within the framework of PB theory the osmotic coefficient turns out to be
given by the expression \cite{Mar55}
\begin{equation}
  \phi_\romPB = \frac{1+\gamma^2}{2\xi},
\end{equation}
where the dimensionless constant of integration $\gamma$ is the
solution of the transcendental equation
\begin{equation}
  \gamma\,\ln\frac{R}{r_0} \; = \; \arctan\frac{1}{\gamma} + 
  \arctan\frac{\xi-1}{\gamma}.
  \label{eq:gamma}
\end{equation}
In the limit of infinite dilution the cell radius $R$ tends to
infinity, which implies $\gamma\rightarrow 0$. Note that for
$\gamma\rightarrow 0$ and $\xi>1$ the right hand side of
Eqn.~(\ref{eq:gamma}) tends to the constant $\pi$, hence the osmotic
coefficient as computed by PB theory (logarithmically) converges to
the well known Manning limiting law $\phi_\infty=1/2\xi$. At {\it
finite} densities it is {\MD always} larger, however.


\section{Beyond the Poisson-Boltzmann solution of the cell model}

The theoretical treatment discussed above suffers from various approximations.
First, the cell model itself is a simplified representation of the
polyelectrolyte solution: It neglects rod-rod-interactions, it is incapable of
describing effects due to the finite {\MD length} of the rods, and it reduces
the solvent to a dielectric continuum. Second, the mean-field approach
discards any inter-ionic correlations which {\MD can} modify the average
charge distribution. In this section we briefly outline a few theoretical
approaches which try to {\MD provide an improved description of the physical
  situation.} Our main purpose is to indicate, in which {\em direction}
changes of the osmotic coefficient are to be expected.


\subsection{Integral equations}

The key approximation in Poisson-Boltzmann theory is the neglect of ionic
correlations by assuming that all $N$-particle distribution functions entirely
factorize. Hence, each ion ``sees'' only a mean ion density but no individual
particles in its vicinity. However, the local ionic order around each ion
influences its energy. Recall for instance that the mean charge density around
an ion in a crystal is zero, but nevertheless the electrostatic energy per ion
is {\em (i)\/} negative, and {\em (ii)\/} the Madelung constant depends on the
crystal structure.

Integral equation theories try to explicitly compute higher order correlation
functions and their impact on the one particle distribution function, i.e.,
the density. The problem of the electric double layer has been treated
frequently in the past for various geometries \cite{KjMa84,Marcelo,Das}. What
concerns us here is that the screening of the macroion by the small
counterions is found to be enhanced. {\MD Since a larger number of ions close
  to the rod implies a smaller number of ions at the cell boundary, this
  effect will {\em lower\/} the osmotic coefficient as compared to
  Poisson-Boltzmann theory.}


\subsection{Correlation corrected density functionals}\label{DHHC}

The correlations neglected in the {\MD mean field} approach can in principle
be recaptured by adding a correction to the free energy functional
corresponding to Poisson-Boltzmann theory \cite{Nor84,PeNo90,Gro91}.  A {\em
  local\/} and {\em stable\/} functional based essentially on a Debye-H\"uckel
solution of the one component plasma has recently been suggested by Barbosa
{\em et al.\/} \cite{BaDe}.  It leads to an enhanced counterion condensation
in the vicinity of the macroion in very good quantitative agreement with
Molecular Dynamics simulations. This again implies a lower density at the
outer cell radius and thus a smaller osmotic coefficient.  The extended free
energy functional translates to a differential equation more complicated than
the PB equation.  However, the equilibrium ion profiles can be conveniently
determined by directly implementing the functional minimization problem along
the lines suggested in Ref.~\cite{Des00}. A comparison of Poisson-Boltzmann
theory, computer simulations and experiments with this correlation corrected
theory will be presented below.


\subsection{Debye-H\"uckel-Bjerrum theory}\label{Bjerrum}

An attempt to go beyond the cell model with the aim of incorporating
rod-rod-interactions has been {\MD suggested} by Kuhn {\em et al.\/}
\cite{KuLe98}.  A solution of charged rods is investigated within the
Debye-H\"uckel-Bjerrum theory, which uses the linearized form of the
Poisson-Boltzmann equation to solve for the charge distributions. The
neglected nonlinearities are approximately recaptured by the explicit
introduction of clusters consisting of a rod and condensed counterions.
Minimizing the free energy of the system, which is constructed as a sum of
electrostatic and entropic terms, gives the cluster distribution and
thermodynamic quantities like the osmotic pressure.  No cell-description is
used, rather, the interaction between individual rods is explicitly computed
and their mutual orientations is averaged over. For the computation of the
screened rod-potential (and at a few other stages) end effects are neglected,
which in the dilute limit entails the Manning limiting behavior of infinite
rods.  For the case of added salt the theory predicts an osmotic coefficient
even below the Manning predictions, but it is unfortunately not
straightforward to extend it to the salt-free case.  A simpler version for the
salt free case \cite{levin97a} that neglects rod-rod interactions, again shows
the osmotic coefficient to be always below the Manning limiting value. This,
and the fact that it decreases upon increasing polyelectrolyte concentration,
raises concerns as to whether the employed approximations lead to a
well-controlled improvement beyond the cell model.


\subsection{Finite rods}\label{finiterods}

An alternative theory beyond the cell model has been {\MD put forward} by
Nyquist {\em et al.\/} \cite{NyHa99}. Here the aim is to explicitly account
for the finite length of the rods. It is argued that in the limit of infinite
dilution a finite rod will not produce Manning condensation and the ions will
become free. This implies the correct limiting osmotic coefficient of $1$
instead of $1/2\xi$ at zero density.  At finite density the condensation is
approximated by a two state model (ions are either free or condensed, and each
rod carries the same fraction of condensed ions). The partition function is
computed in a random phase approximation, which finally gives the equilibrium
condensed fraction as a function of density. In the dilute regime this
fraction is found to {\em increase\/} with density and the osmotic coefficient
is correspondingly found to {\em decrease}.  {\MD Furthermore}, the {\MD
  calculated} osmotic coefficient is larger than the Poisson-Boltzmann result.
This is in strong contrast to any solution of the problem which assumes
infinite rods.

Up to now there is no systematic comparison of theory and experiment, however,
the model seems to overestimate considerably the osmotic coefficient when
compared to experimental data \cite{AuAl69,BlWi}.  Possible reasons for this
shortcoming may be related to the two state model or the random phase
approximation employed in Ref. \cite{NyHa99}.


\section{Simulations}

After restricting to the simplified models mentioned above, simulations can be
very successfully used to test the available theories, demarcate their range
of applicability, and indicate the trends of deviations which are to be
expected. Consequently, versions of the cylindrical cell model have been used
several times in the past as a basis for simulations
\cite{DeHo00a,Gul89,NiGu91,LyNo97}. An important observation has been that
condensation is generally enhanced compared to Poisson-Boltzmann theory, which
is {\MD frequently} attributed to the correlations neglected by the latter.

In this article we use molecular dynamics simulations to determine the osmotic
coefficient for a cell model, the parameters of which have been mapped to a
specific polyelectrolyte: poly(para-phenylene) (PPP, see Fig.~1 and below).
We take a cubic simulation box of length $L_\romb$, one charged rod parallel
to an edge, and the correct number of monovalent counterions which leave the
system electrostatically neutral. The ions interact via a purely repulsive
Lennard-Jones potential \cite{ste95a}, giving them a diameter of
$4.4\,\text{\AA}$.  For the rod-ion-interaction a similar potential was used,
in which {\MD the hard core was shifted to larger radii, such that the
  distance of closest approach between an ion and the rod was
  $7\,\text{\AA}$.}  This value has been inferred from an analysis of small
angle (neutron or X-ray) scattering experiments on PPP
\cite{KaJe87,MaKa98,GuBl00}. Each ion carries a negative unit charge in its
center and the rod carries a sequence of positive unit charges along its axis
at a distance of $2.15\,\text{\AA}$.  Together with a Bjerrum length of
$7.31\,\text{\AA}$ (corresponding to water at $40\,^{^\circ}\!\text{C}$) this
gives a charge parameter $\xi=3.4$. After switching on periodic boundary
conditions this geometry yields an infinite array of infinitely long rods
sitting on a square lattice. {\MD When comparing to the cylindrical cell model
  we will map the rectangular cell belonging to one such rod with a
  cylindrical cell of radius $R=L_\romb/\sqrt{\pi}$ \cite{squareworries}.}

The electrostatic interactions in this periodic boundary geometry were
computed with the help of P$^3$M routines \cite{DeHo98}, and a Langevin
thermostat \cite{GrKr86} combined with a velocity-Verlet-integrator
\cite{AlTi97} (with time-step 0.01 in Lennard-Jones units) was implemented to
drive the system into the canonical state. The saturation of the electrostatic
energy was used to test for equilibration. A more detailed description can be
found in \cite{DeHo00b}. We simulated the rods at a monovalent counterion
concentration of 2.18, 6.79, and 10.2 mmol/l, with a number of 4.5, 1.8, and
1.2 million MD steps, respectively. The corresponding number of monovalent
ions in the simulation cell were 277, 1256, and 1024.  The osmotic pressure
was taken as the average of the $xx$- and $yy$-component of the stress tensor,
given that the rod points along the $z$-axis.


\section{Experiment}

From its derivation it is obvious that the cylindrical cell-model can be valid
only if the macroions are sufficiently stretched. DNA in salt-free solution
can be regarded as such a model polyelectrolyte \cite{AuAl69,MaKa98,RaCo00}.
However, the stability of the helical conformation is severely impeded if the
ionic strength is low, and measurements in salt-free solution must be carried
out at low temperature to avoid the melting of the helix. To the author's best
knowledge the study of Auer and Alexandrowicz \cite{AuAl69} has been the only
one in which strictly salt-free solutions of DNA have been studied. A more
recent investigation employed solutions of DNA with additional $2\,\text{mM}$
{\MD and $10\,\text{mM}$} added salt \cite{RaCo00}. However, in this case
counterions and salt ions will establish a Donnan equilibrium
\cite{Oos71,Kat71,Don24}, which results in a drastic reduction of the
small-ion contribution to the pressure. Its dominant origin at low
polyelectrolyte density is not yet fully understood \cite{saltcomment}.

Recently, an investigation of the osmotic coefficient of a synthetic rod-like
polyelectrolyte has been presented \cite{BlWi}. The system studied there
consists of a poly(p-phenylene) (PPP) backbone to which positively charged
side-chains have been attached (see Fig.~1). The fully aromatic backbone
exhibits a high stiffness with a persistence length of approximately
$22\,\text{nm}$ \cite{Gal94}.  The number-average contour length of
PPP-macroions {\MD used in the experiment} is of the same order. Hence, the
PPP-polyelectrolytes may be treated as rod-like in {\MD very good}
approximation.  The system has a charge parameter $\xi=3.4$ of similar
magnitude as DNA ($\xi=4.2$). The excellent chemical stability of the
PPP-chains allows to study these polyelectrolytes in salt free solutions
without any problem.

In Ref. \cite{BlWi} the osmotic coefficient of the PPP system was measured for
two kind of counterions, iodine and chlorine. Both sets of experimental data,
which have been obtained in virtually salt free solution, clearly indicate that
the Poisson-Boltzmann solution of the cell-model makes a fairly good
prediction for the osmotic coefficient, but systematically overestimates it
upon closer look. This is also in agreement with the earlier results obtained
on DNA by Auer and Alexandrowicz \cite{AuAl69}. Moreover, it became apparent
in this study that there are specific interactions of the counterions with the
macroion, which can be as large as the differences from the Poisson-Boltzmann
solution itself. This was concluded from the fact that chlorine counterions
lead to a considerably smaller $\phi$ than iodine counterions. Such specific
interactions have been found for many flexible polyelectrolytes and
extensively discussed in the earlier literature \cite{Man88,Kat71}.  The
errors in the experiment are hard to estimate, but are on the order of 7\%.

\section{Comparison of theory, simulations and experiment}

It has to be understood that an analysis of the osmotic coefficient $\phi$
measured in salt free solutions within the framework of the Poisson-Boltzmann
solution of the cylindrical cell model proceeds virtually without adjustable
parameters. Here, $\phi$ is solely determined by the charge parameter $\xi$,
which is fixed by chemistry, the rod radius $r_0$, which has been measured
experimentally, and the polyelectrolyte concentration, which is fixed by the
experiment. The latter parameter determines the cell radius $R$ (see the
discussion in Ref.~ \cite{BlWi}). For the theoretical or simulational modeling
those values were taken as input parameters and were not adjusted such as to
fit the measured data.
 
Fig.~\ref{fig:phi_PPP2I} summarizes the results. It shows the osmotic
coefficient $\phi$ of a PPP solution as a function of counterion concentration
$c_\romc$ as predicted by Poisson-Boltzmann theory, the DHHC
correlation-corrected treatment from Sec.~\ref{DHHC}, Molecular Dynamics
simulations and experiment.
  
To begin with, it is important to note the fine vertical scale.
Poisson-Boltzmann theory predicts $\phi$ to be smaller than 1 and vary roughly
within the range $0.18 \ldots 0.22$.  The measured values accumulate around
$0.18$ (iodine) and $0.16$ (chlorine). Hence, the dominant change in $\phi$, a
reduction by a factor of $5$, is correctly accounted for. However, on the
enlarged scale of Fig.~\ref{fig:phi_PPP2I} it is visible that the measured
values are systematically lower than the prediction, although still higher
than the Manning limit $1/2\xi$ of infinite dilution.

Both the correlation-corrected DHHC theory as well as the simulations, which
in principle capture all kinds of ion correlations, show a decrease in the
osmotic coefficient. Since these two totally different approaches agree so
well, we believe that they indeed give a good description of the influence of
correlations. However, they do not lower the osmotic coefficient sufficiently
enough as to agree with the experimental results.

The deviation from the Poisson-Boltzmann curve increases for higher densities,
which is true for the DHHC and simulational part as well as for the
experiment. This appears plausible if one recalls that correlations become
more important at higher densities.  Unfortunately, the increasing systematic
experimental errors in this regime are very difficult to quantify.  Due to
this fact, we give no significance to the apparent slight decrease of $\phi$
for the I$^-$ ions.

\section{Discussion}

The fact that Poisson-Boltzmann theory overestimates the osmotic coefficient
as been observed previously.  Careful studies of typical flexible
polyelectrolytes in solution (see Ref.~\cite{Man88,Kat71} and further
references given there) indicated that agreement of the Poisson-Boltzmann cell
model and experimental data could only be approached if the charge parameter
$\xi$ was renormalized to a higher value. The motivation given for this ad hoc
modification was the assumption of a locally helical or wiggly main chain.
Hence, the counterions ``see'' more charges per unit length, i.e., a macroion
having a higher charge parameter.  However, the results obtained for
stiff-chain macroions \cite{BlWi} show that the osmotic coefficient is lower
than the Poisson-Boltzmann results even for systems where the local
conformation of the macroion is absolutely rod-like, therefore this
explanation can not be used.  Also, the measured $\phi$ as a function of
density has a functional form different from the Poisson-Boltzmann prediction,
and {\em no\/} value of $\xi$ gives a curve that fits all data points.

Since Poisson-Boltzmann theory neglects all ion-ion correlations, it is
tempting to assume that their incorporation into the theoretical treatment
would resolve the discrepancy. However, our comparison shows that these
correlational effects can be made responsible only for part of the deviations.
Since the different approaches using a correlation-corrected density
functional theory and Molecular Dynamics simulations agree very well with each
other, we believe that the discrepancy between them and the experiment is not
due to theoretical or simulational errors, but has to be looked for in the
approximations underlying the cell model itself.

If there were any excess salt present in our system, the osmotic coefficient
would be reduced \cite{Kat71,RaCo00,podgornik00}. However, we performed
numerical solutions of the Poisson-Boltzmann equation, taking into account
various amounts of excess salt as well as the constraints due to the Donnan
equilibrium, which showed two things (see Fig.~\ref{fig:phi_PPP2I}): First,
the functional form of $\phi(c_\romc)$ is not compatible with a roughly
constant value of the osmotic coefficient. Rather, it approaches the salt-free
result at high $c_\romc$ and drops strongly (below the Manning limit) at low
$c_\romc$.  Second, our measurements at small counterion concentration are
incompatible with the assumption of an excess salt concentration exceeding a
few micromolar. In fact, the strong decrease of the osmotic coefficient at low
concentrations, even below the Manning limiting value for the Cl$^-$ ions, can
be explained by different very small excess salt concentrations for both
experimental sequences (2 mikromol for I$^-$, 20 mikromol for Cl$^-$, compare
Fig.~2).  However, at our high concentrations, where the discrepancy is most
apparent, any remaining excess salt is least relevant, and hence cannot
explain neither the difference in the osmotic coefficient between Cl$^-$ and
I$^-$ ions nor the decrease of $\phi$ compared to the PB prediction.

One obvious approximation of the cell model is the neglect of end effects of
the rods by assuming them to be infinitely long. However, as we have discussed
in Sec.~\ref{finiterods}, correcting this point would lead to an {\em
  increased} osmotic coefficient, since finite rods show less condensation
(indeed, none in the limit of zero density). Moreover, the theoretical
treatment of finite rods by Nyquist {\em et al.\/} \cite{NyHa99} produces too
large coefficients even at finite densities. Hence, finite size effects of the
rods are unable to explain the additional reduction of the osmotic
coefficient.

The theoretical approach mentioned in Sec.~\ref{Bjerrum} yields a reduced
osmotic coefficient. However, it is unable to predict values above the Manning
limit, and shows a {\it decrease} of $\phi$ with increasing concentration,
contrary to the experimental indications.  The slight apparent decrease for
the iodine is, if the errors are considered, not significant.

The remaining deficiency of the cell model is the neglect of any molecular
detail, in particular: The solvent is described by a dielectric continuum.
However, hydration effects are well known to be important in many
circumstances \cite{CoAy99,Wi97,Bi94} and to depend on the particular objects
to be hydrated \cite{Bi94}. This of course applies to the counterions as well
as to the charged groups of the macroion. Indeed, the difference between the
osmotic coefficient measured for chlorine and iodine ions indicates that they
are not just charged spheres.  Similarly, the macroion is not just a charged
rod.

Unfortunately, those last problems with the cell model are the hardest to deal
with theoretically. Even predicting the direction into which changes are to be
expected are very complicated \cite{huenenberger99a}. Obviously, much work
still has to be done in order to arrive at an improved quantitative prediction
of the osmotic coefficient of rodlike polyelectrolytes, since all the
``obvious'' modifications will not suffice, as we tried to show.


\section{Conclusion}

We tested the ability of the cell-model to explain the measured osmotic
coefficient of solutions containing rodlike polyelectrolytes and monovalent
counterions. We compared the Poisson-Boltzmann solution of the cell model, an
improved local density approximation (DHHC), and simulational results of the
model to recent experimental data.

Our findings are that the Poisson-Boltzmann solution of the cell model
systematically overestimates the osmotic pressure. We showed that only a part
of this discrepancy between theory and experiment is due to the neglect of
correlations in the mean field approach, which will lower the osmotic
pressure. We argued that other simple explanations like a locally curved
main-chain conformation or the neglect of remaining excess salt do not apply
here. We also outlined why dropping the assumption of infinite rods would in
fact worsen the disagreement

We are led to the conclusion that the remaining discrepancy between experiment
and a theoretical description of the system should be looked for on the level
of molecular detail. This includes a better description of the solvent,
hydration effects, or van der Waals forces. We are well aware of the fact that
this requires formidable theoretical or numerical efforts, but we would like
to stress that a full agreement between theory and experiment cannot and
should not be expected on the level of a restricted primitive model for
electrolytes, even if all correlations are properly taken into account. We
believe that our comparison provides some evidence to this statement and
thereby hope to motivate future theoretical and experimental work to resolve
these issues.


\section*{Acknowledgments}

The authors gratefully acknowledge financial support by the Deutsche
Forschungsgemeinschaft, Schwerpunkt Polyelektrolyte, and a computer time grant
hkf06 from NIC J\"{u}lich. We furthermore enjoyed discussions with M. Barbosa,
P.L. Hansen, and R. Podgornik.





\clearpage

\begin{figure}\label{fig:ppp}
  \begin{center}
\epsfig{file=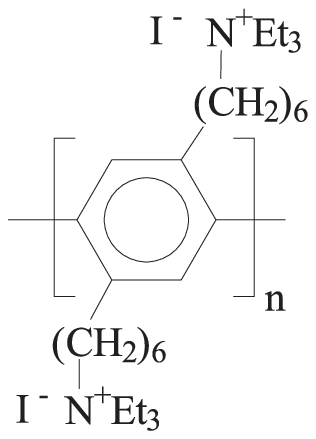,width=5cm}
  \end{center}
  \vspace*{2cm}
  \caption{Constitution formula for poly(p-phenylene). The fully aromatic
    backbone exhibits an excellent chemical stability. Its uncharged precursor
    has a persistence length of approximately $22$ nm. The degree of
    polymerization used in the studies in Ref.\cite{BlWi,GuBl00} was located
    between 20 and 40. Therefore the contour length equals approximately one
    persistence length at most.}
\end{figure}


\clearpage

\begin{figure}\label{fig:phi_PPP2I}
  \begin{minipage}{\textwidth}
    \centering \input{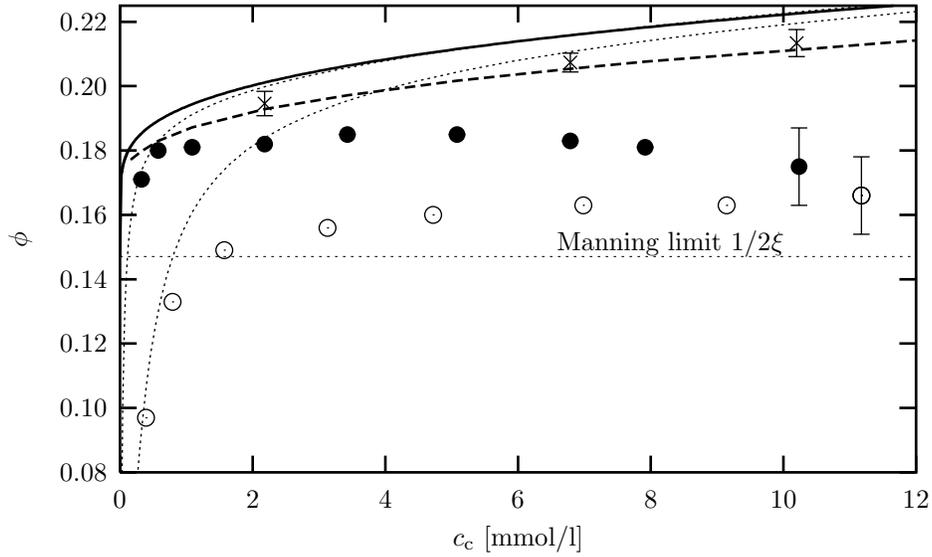} 
  \end{minipage} 
  \vspace*{5cm}
  \caption{Osmotic coefficient as a function of counterion concentration
    $c_\romc$ for the poly(p-phenylene) systems described in the text.  The
    solid curve is the PB prediction of the cylindrical cell-model, the dashed
    curve is due to the correlation corrected PB theory from Ref.~\cite{BaDe}
    (see Sec.~\ref{DHHC}). The circles are experimental results using iodine
    (full) and chlorine (empty) counterions. Errors are of the order of 7\%
    for all values, but only indicated for one measurement. The crosses
    originate from the MD simulations described in the text.  Furthermore, the
    thin dashed lines are PB predictions assuming excess salt concentrations
    of 2 micromolar (upper) and 20 micromolar (lower), which are to be
    understood as crude estimates of the actual excess salt content. The
    Manning limiting value of $1/2\xi$ is also indicated as a horizontal
    dashed line.}
\end{figure}



\begin{thebibliography}{99}
\bibitem{squareworries} One may worry about impacts of the non-cylindrical
geometry on the simulation. However, in the strongly charged regime
most ions reside in a fairly close vicinity of the rod, where the
potential is still cylindrical. It is those ions that will actually
feel correlation effects and thereby trigger a readjustment of the
whole ion profile as compared to Poisson-Boltzmann theory.
Furthermore, the cell model partly relies on the assumption that the
actual shape of the Wigner-Seitz-cell does not matter.
\bibitem{saltcomment} Usually the contribution to the total osmotic pressure
coming from the counterions by far exceeds the macroion part. However,
this may no longer hold if the ionic contribution almost cancels at
low polyelectrolyte concentration due to the Donnan effect, and the
macroion may reappear. Ref. \cite{RaCo00} speculates along similar
lines.
\end{thebibliography}
\end{document}